# Detecting and Classifying Android Malware using Static Analysis along with Creator Information

Hyunjae Kang[1], Jae-wook Jang[1], Aziz Mohaisen[2], and Huy Kang Kim[1]

[1] Korea University, Seoul, Korea
`janetk1004@gmail.com,{changkr, cenda}@korea.ac.kr`
[2] Verisign Labs, Reston, VA 20190, USA
`amohaisen@gmail.com`

**Abstract.** Thousands of malicious applications targeting mobile devices, including the popular Android platform, are created every day. A large number of those applications are created by a small number of professional underground actors, however previous studies overlooked such information as a feature in detecting and classifying malware, and in attributing malware to creators. Guided by this insight, we propose a method to improve on the performance of Android malware detection by incorporating the creator's information as a feature and classify malicious applications into similar groups. We developed a system that implements this method in practice. Our system enables fast detection of malware by using creator information such as serial number of certificate. Additionally, it analyzes malicious behaviors and permissions to increase detection accuracy. The system also can classify malware based on similarity scoring. Finally, we showed detection and classification performance with 98% and 90% accuracy respectively.

**Keywords:** Mobile malware, Android security, malware detection, malware classification, creator information

## 1 Introduction

Mobile devices, especially smartphones, made a great contribution toward environment of fast and massive information sharing. Smartphone market is expanding consistently every year, enriching our digital lives in the sense of communication and entertainment. However, the increase of usage of mobile device caused a serious problem at the same time. McAfee collected 2.47 million new mobile malware samples and total 3.73 million malware samples in 2013. The malware totaled at the end of 2013 increased by nearly 200% from the end of 2012 [1].

Mobile malware mainly targets Android platform. The reason is that Android user occupies the largest part of the mobile market, and unfortunately, the platform allows easy distribution of malware created or repackaged as well. There were 827 new families or variants of mobile malware collected in 2013 by F-Secure. They reported that most of the families were based on Android platform; exclusive of 23 families only [2]. Also, 275 new Android threats (new families or new variants) with only two threats of other mobile platforms have collected by them [15].

In order to mitigate the threat on mobile device, various efforts have been made to detect and analyze mobile malware. There are two main approaches, static analysis and dynamic analysis. Many studies of static analysis use requested permission to check the risk of an application, but the methods of only permission-based have its limits to detect malware accurately. There are also other detection methods analyzing other features such as API instead of permission. Different from static analysis, dynamic analysis usually deals with features like dynamic code loading and system call that can be collected while an application is running.

While process of signing the application is required before distribution, a certificate may be a clue to trace the developer. Applications are often attributed to their developers using certificates tied to the developer's signing credentials, although investigating certificate's information has not been actively studied yet. Very recently, Jerome et al. [23] used certificates signing malware to isolate applications as a potential malware set, which showed the possibility of filtering malware with a predefined list of malicious certificates. We hypothesized that certain developers have a significant role in the creation and distribution of malicious apps. Indeed, we found that only 4% of total certificates collected from malware are signing as much as 70% of the malware samples discussed in this paper.

Our proposed system provides a light-weight malware detector using serial numbers of certificates and similarity-based malware classifier. To this end, the contributions of this work are as follows:

- Our system contains malware detection method using static analysis associated with creator information. The system uses serial numbers of certificates, getting light-weighted process to detect malware. Also, it analyzes particular parts of applications based on functionality and permission, and achieved high detection accuracy.
- Based on similarity scoring algorithm, our system classifies malware into same expected families. The algorithm compares API sequence, permissions, and system commands extracted from malware. Grouping similar malware samples and demonstrating signature of each group can help to figure out a number of samples.
- We automated our proposed system and experimented on collected real samples; 51,179 benign applications and 4,554 malware samples. It determined most of the applications correctly only except 454 applications (less than 1% of all samples). Then, the system classified malware families by gathering them with 90% accuracy tested on the applications decided as malware from preceding detection process.
- We analyzed distributions of certificates signing applications of our dataset samples. Benign application sets and malware sets are compared based on the number of applications and malware families created by each certificate.

## 2 Related Works

There are mainly two approaches to analyze Android malware: static analysis and dynamic analysis. Static analysis is a way to check functionalities and maliciousness of an application by analyzing its source code, without executing the application. It is useful for finding malicious behaviors that may not operate until the particular condition occurs. On the other hand, dynamic analysis is a method to examine an application during runtime. It may miss some parts of the code that not executed, but it can easily reveal certain malicious behaviors too complicated to find by static analysis.

**Table 1.** Related works on Android malware analysis

| Method | Main Features | Papers |
|---|---|---|
| Static analysis | Requested permission | [4], [6], [12], [13] |
| | Requested permission, Android component | [3] |
| Dynamic analysis | Presence of root exploits, use of encryption, dynamic code loading | [7] |
| | API call, system call | [9] |
| | System call, system log | [14] |
| | System call | [11] |
| Hybrid analysis | Requested permission, behavioral footprint, dynamic code loading | [5] |
| | API call sequence, dynamic code loading | [8] |
| | Requested permission, intent, native API call | [10] |

Analyzing the requested permissions is one of the popular ways in static analysis. An application asks users for permissions before installation, so it notifies to the user what information and resources the application can access. Enck et al. [3] proposes a security service system, named Kirin, which certificates an application at the moment of installing, using a set of predefined security rules. They analyze the type of malicious behaviors and defined the rules configured with permission and intent information. AdDroid system separates privilege from advertising framework to Android platform [4]. It prevents advertising library to access sensitive information allowed for other permission of the application. Felt et al. [6] analyze real-world mobile malware from 2009 to 2011. Also, they discuss effectiveness of using permission distribution in order to classify benign application and malware. Peng et al. [12] applies simple Naïve Bayes to requested permissions, and develop hierarchical mixture model. Also, Sarma et al. [13] compares distribution of permissions between benign application and malware, then determine critical permissions to detect malware. Both [12], [13] studies propose a risk-scoring method for applications using their own algorithms. Permission based malware detection model's major weakness is low accuracy. The problem arises from the laxity of Android permission architecture. Application developers can request permissions that are not necessary; the distribution of permissions becomes incorrect.

On the other hand, there are dynamic analysis based approaches. RiskRanker is an automated system that detects particular malicious behaviors [7]. First, it checks whether native code of an application contains the known exploit codes or not. Secondly, it captures certain behaviors like encryption or dynamic code loading. AppsPlayground performs functions like information leakage detection, sensitive API monitoring, and kernel level monitoring [9]. Burguera et al. [11] propose a system named Crowdroid that monitors and logs system call and send it to a central server. At the server side, the system operates K-means clustering to classify benign app and malware. Jang et al. [14] analyze integrated system logs including system calls, generated while malware running on an emulator. Their system, Andro-profiler, makes a behavior profile with analyzed system logs in human-readable form. By comparing it with the predefined profiles of malware families, it can detect and classify malware with high accuracy.

There are also hybrid approaches that adopt both static analysis and dynamic analysis. Zhou et al. [5] detect malware samples of known families with permission based behavioral footprinting scheme. They manually analyze and abstract essential malicious behaviors as footprints to complement permission based analysis. The paper also includes heuristic based filtering scheme to detect samples of unknown malicious families, which mainly monitors dynamic code loading. Another paper that applied both analyses explains previous signature based approaches are not effective since a signature like cryptographic hash can be easily changed by malware developer [8]. They extract API call sequence and check dynamic downloading of malicious code, then generate three levels of signatures. These signatures facilitate identifying malicious code segment, along with class association among applications. Spreitzenbarth et al. [10] presents an automated analyzing system, Mobile-Sandbox, combines static and dynamic analysis. It parses an application's permission and intent information and analyzes the suspiciousness of them. Then, it performs dynamic analysis to log actions especially those based on native API calls.

## 3    Proposed System for Detecting and Classifying Android Malware

### 3.1    Android platform and malware

#### 3.1.1    Features of Android applications

In this section, we explain features of Android platform we mainly used in profiling applications. The Android OS has a distinctive structure. Applications on Android do not have a unique entry that programs usually have on other OS. They are made up of Android components; activity, service, broadcast receiver, and content provider. Activity is a UI component related to the screen while service is a background process which is invisible to the user. Broadcast receiver waits for the signals from the system and wakes up the proper actions after receiving. Content provider plays a role of an intermediate unit to share data between applications. These four components work individually, therefore a unit for delivering messages is needed; namely intent. Intent transfers from activity to activity, containing specific instructions about what the application wants.

API (Application Programming Interface), documented in Android SDK, is a set of functions provided to control principal actions of Android OS. It is much efficient to consider certain APIs often used by malware, rather than extracting all APIs from the source code of an application. Seo et al. [17] analyzed malware samples and determined suspicious APIs often used by malware. They listed suspicious APIs and compared the number of use between malware and benign applications. We manually collected additional APIs by checking all APIs in Android SDK, which operate in a way similar to the suspicious APIs defined in [17]. These APIs are related to functions such as collecting the user or device information, accessing websites, sending and deleting SMS, and installing an application.

122 permissions are provided in Android platform to inform the user what actions will be performed and which resources will be accessed by an application. Sarma et al. [13] compared two dataset, applications from the Android Market and malicious applications. They analyzed the distribution of permissions requested by each dataset and found 26 risky permissions to the security and privacy. Peng et al. [12] used the critical permissions applying Naive Bayes Models to score risk of an application. We use those permissions in our system. One of the permissions, INTERNET, is excluded, since this permission is required for most applications. We regarded that relatively small difference between the percent of applications requested from Android market applications and malware will not play an important role in detection and classification. INSTALL_PACKAGES, a permission usually used for installing a new package downloaded from a server, is included instead. While requested permissions are notified before download, there is another way to extract permissions by analyzing call-graph. Au et al. [18] specified the list of permissions required by every API call, and provided the permission mappings. In our system, we extracted 26 critical permissions applying PScout mapping, along with requested permission. We named this feature as API-related permission.

When distributing an application, the creator signs it by his private key and a standard certificate of the public key is generated. There are blanks for the creator's name, organization, and location, while generating a certificate. However, the creator can fill them up with false information since the process has no steps of confirmation. The certificate has a unique serial number according to RFC 2459, the X.509 standard. Based on that, one can check whether certificates are the same or not by comparing the serial number. We hypothesize that there may be frequently-detected serial numbers in various malware variants, and analyzed the results in this paper.

**3.1.2    Threats from Android malware**

Malware can infect Android devices via many infection routes; as a downloaded application from visits to malicious websites, spam, malicious SMS messages, and malware-bearing advertisements [1]. After malware infects a target device, behaviors of the malware can be categorized depending on their purpose. Zhou and Jiang [16] classified malicious behaviors into privilege escalation, remote control, financial charge, and information collection. Also, Seo et al. [17] listed monetization, information stealing, mobile botnet, and root privilege acquisition as categories. Through examining these behaviors, we rearranged those to particular functions to use in our malware detector module; the functions are using system commands on root privilege, causing financial fraud by hiding SMS notification from the user, and collecting sensitive information.

### 3.2 Detecting and classifying malware

We designed a system for detecting and classifying Android malware. Fig. 1 shows the overall architecture of the system. It is composed of three parts; parser module, malware detector module, and malware classifier module. First, the parser module extracts strings like certificate information, APIs, permissions, commands, and intents from an application. Then, the malware detector module checks whether a certificate serial number of the application is included in a predefined serial number blacklist. If not, it additionally analyzes the application based on other features like APIs, intents, system commands, and permissions. Finally, the malware classifier module groups malware with same families by comparing the similarity of API sequence, permissions, and system commands between the applications. The parsed data and outputs of detector and classifier modules are updated in a shared database.

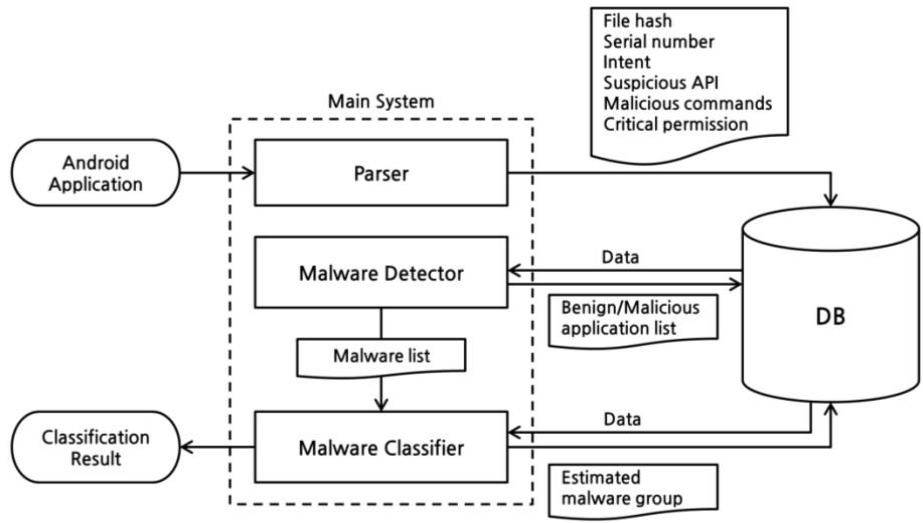

**Fig. 1.** Overview of proposed system

We collected malware samples from malware repository websites such as VirusShare [19], Contagio Mobile [20], and Malware.lu [21]. The samples are collected during January to August 2013. We used malware description of F-Secure antivirus to tag representative family association and labels. Based on the F-Secure antivirus description, malware families that include too many samples were cut down, and families that have only few samples were excluded. The number of refined malware samples is 4,554. Also, we collected 51,179 applications during the same period by downloading from the Android market, GooglePlay, and assumed that they are benign.[*]

### 3.2.1 Serial number blacklist

Among the collected malware samples, we extracted a serial number of each sample's certificate. 622 serial numbers were observed in total malware samples, but only 24 unique serial numbers comprised 70% of the samples, and surprisingly, 50% of malware samples were signed by five serial numbers. That means particular serial numbers are frequently used by malware creators. However, the counted number of malware signed by the same certificate serial number may have an error, because the counted number depends on the collected samples. In addition, we analyzed commonly found serial numbers in the various malware families and their variants. Among the 622 serial numbers, there were 137 numbers that generated more than 2 families or variants of malware. There were 485 numbers that generate only one kind of families or variants. Rest of serial numbers used to create many malware families or variants, and the number of families or variants varied from 2 to 11.

We made a blacklist of serial numbers from the resulting certificates. The serial numbers creating only one type of malware were excluded from the blacklist. The serial numbers found in over two malware families or variants were included. '93:6e:ac:be:07:f2:01:df' was deleted in the list, since it is a serial number of standard test key for native applications that are built in a device or an emulator. As a result, the serial number blacklist had 136 numbers for malware detection.

### 3.2.2 Likelihood ratio of permission

Benign applications and malware groups have different tendencies of requesting permissions. Malware can request more permissions than benign applications, or can often request permissions that have risks related to the privacy problem or financial fraud. We analyzed the distribution of the critical permissions in each benign application sample and malware samples we have, to calculate likelihoods of the permission.

**Table 2.** Critical permission distribution in benign applications and malware samples (%)

| Permission | Requested | | API-related | |
|---|---|---|---|---|
| | *Benign app* | *Malware* | *Benign app* | *Malware* |

---

[*] Our dataset is available at http://ocslab.hksecurity.net/andro-tracker

| Permission | Requested | | API-related | |
|---|---|---|---|---|
| | Benign app | Malware | Benign app | Malware |
| ACCESS_COARSE_LOCATION | 16.61 | 53.78 | 20.52 | 56.28 |
| ACCESS_FINE_LOCATION | 16.96 | 51.89 | 17.72 | 55.82 |
| CALL_PHONE | 6.57 | 26.92 | 0 | 0 |
| INSTALL_PACKAGES | 0.32 | 12.67 | 0 | 0 |
| PROCESS_OUTGOING_CALLS | 0.63 | 1.80 | 0 | 0 |
| READ_CONTACTS | 5.82 | 24.95 | 1.72 | 0.22 |
| READ_SMS | 1.22 | 27.82 | 0 | 0 |
| SEND_SMS | 1.82 | 43.98 | 1.04 | 35.07 |
| WRITE_CONTACTS | 2.08 | 1.47 | 1.72 | 0.22 |
| BLUETOOTH | 1.51 | 4.04 | 1.21 | 2.37 |
| BLUETOOTH_ADMIN | 1.21 | 2.77 | 0.95 | 0.53 |
| GET_ACCOUNTS | 4.40 | 4.90 | 3.39 | 3.67 |
| MOUNT_UNMOUNT_FILESYSTEMS | 0.80 | 20.62 | 0 | 0 |
| NFC | 0.26 | 0.04 | 0.15 | 0 |
| READ_CALENDAR | 0.97 | 0.04 | 0 | 0 |
| READ_HISTORY_BOOKMARKS | 0.93 | 7.88 | 0.25 | 5.64 |
| READ_LOGS | 1.39 | 28.59 | 0 | 0 |
| READ_PHONE_STATE | 24.10 | 96.55 | 12.00 | 69.19 |
| RECEIVE_MMS | 0.20 | 1.05 | 0 | 0 |
| RECEIVE_SMS | 1.66 | 37.66 | 0 | 0 |
| RECEIVE_WAP_PUSH | 0.05 | 3.01 | 0 | 0 |
| RECORD_AUDIO | 3.13 | 22.20 | 2.53 | 27.84 |
| WRITE_CALENDAR | 0.85 | 0 | 0 | 0 |
| WRITE_EXTERNAL_STORAGE | 32.25 | 82.50 | 0.10 | 0.68 |
| WRITE_HISTORY_BOOKMARKS | 0.57 | 7.07 | 0.04 | 0.02 |
| WRITE_SMS | 0.77 | 5.67 | 0 | 0 |

Table 2 shows the ratios of the permissions for each category, requested and API-related, by benign samples and malicious samples. The likelihood of permission on each category can be calculated by Naïve Bayes Classifier. The permissions should be relatively independent to multiply each probability of permission. Au et al. [18] found out that most of the Android permissions have subtle correlation with any other permission. They recognized only 15 pairs of permissions have small dependency in all permission set. The critical permissions we used are not included in these pairs so we can assume that they are relatively independent.

Let n and m be number of applications and number of critical permissions respectively. The permission vector for application $i$ is $a_i = (a_{i,1}, a_{i,2}, \cdots, a_{i,m})$, where

$$a_{i,j} = \begin{cases} 1, & \text{if application } i \text{ uses permissoin } j \\ 0, & \text{otherwise} \end{cases}.$$

Also, we put $c_i \in \{benign, malicious\}$ which indicates category of the application $i$. Then,

$$P(c_i|a_i) = P(c_i|a_{i,1}, a_{i,2}, \cdots, a_{i,m}) = \prod_{j=1}^{m} P(c_i|a_{i,j}).$$

Using Bayes' Theorem, the conditional probability of category $c_i$ given variable $a_{i,j}$ which informs the usage of permission can be written as

$$P(c_i|a_{i,j}) = \frac{P(a_{i,j}|c_i) \cdot P(c_i)}{P(a_{i,j})}.$$

Then, ratio of probabilities is

$$\frac{P(malicious|a_{i,j})}{P(benign|a_{i,j})} = \frac{P(a_{i,j}|malicious) \cdot P(malicious)}{P(a_{i,j}|benign) \cdot P(benign)}.$$

We assume $P(c_i = malicious) = P(c_i = benign)$; it is a situation of having no information about the category of the application, so the application is supposed to have a variable of any category value following uniform distribution. By multiplying probabilities of $m$ permissions, the likelihood ratio $\Lambda$ is

$$\Lambda(a_i) = \frac{P(c_i = malicious|a_i)}{P(c_i = benign|a_i)} = \prod_{j=1}^{m} \frac{P(a_{i,j}|c_i = malicious)}{P(a_{i,j}|c_i = benign)}.$$

Malware can be detected by comparing likelihood ratio with some predefined threshold value $T_L$.

If one of the conditional probabilities is zero, then the whole multiplication becomes zero. To avoid such case, the conditional probabilities are calculated using the Laplace estimation,

$$P(a_{i,j}|c_i) = \frac{\sum_{i=1}^{n} a_{i,j} + 1}{n + 2}.$$

### 3.2.3 Malware detection

In the previous section, we defined the representative malicious behaviors using the root privileged system commands, concealing SMS notification to charge rate, and so forth. These kinds of well-known malicious behaviors can be detected by our system. It uses the serial number of certificates as one of the features to detect Android malware. Also, permission based rule is applied as one of the detection rules. Our detection algorithm is designed with mainly three ideas: 1. existence of certain serial numbers generating many kinds of malware, 2. malicious behaviors like command usage on root privilege, hiding SMS notification, and collecting sensitive information, and 3. high likelihood of malware under given distribution of permissions.

The detection algorithm starts by checking the application's serial number of a certificate for fast scanning by applying the blacklist we established in the previous section. There were a few applications that have serial number in the blacklist but do not use any suspicious APIs. The step excludes them to avoid over-detecting.

Secondly, the algorithm checks the usage of the system commands. Seo et al. [17] listed commonly used commands by malware. We listed them by excluding the commands used only in a few malware samples. 'chmod', 'insmod', 'su', 'mount', 'sh', 'killall', 'reboot', 'mkdir', 'getprop', 'ln', and 'ps' are commands often used by malware, and they run on rooted Android device. If any of those strings is found in the source code of an application, we mark it as malware. The commands are executed after the malware obtains root privilege on the device. Also, our list contains 'gingerbreak' and 'rageagainstthecage' because they are root exploits. These commands, which run on rooted device or to root a device, are found in malicious codes.

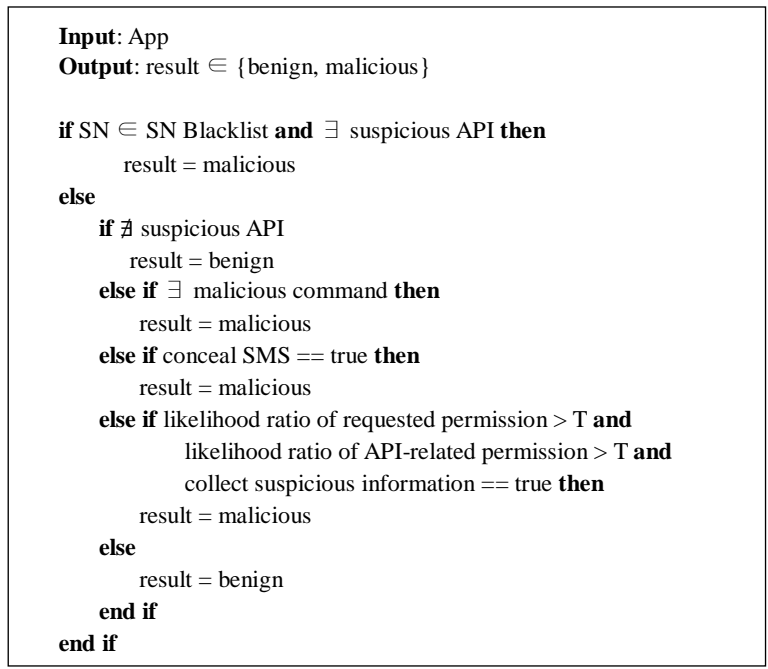

**Fig. 2.** Detection algorithm

The next step is finding malware that conceals SMS notification. Malware that behaves in this way has a purpose of subscribing to premium services confirming and noticing by SMS. These applications use sendTextMessage() to send SMS. Also, we checked specific intent information to detect malware which conceals SMS notification. Some malware receive notification of SMS with the highest priority, and then make the message not delivered to other applications. They get the highest priority of SMS receiving intent and call abortBroadcast() to hide a notification of SMS to other applications and users. This malicious behavior can be found by searching the intents. The step checks whether an application uses above methods and an intent filter, to catch such malware.

For the final step, the algorithm adopts a permission based detection rule. It calculates likelihood ratio under given distribution of permissions. Two likelihood ratios are obtained using requested critical permissions and API-related critical permissions. To complement a limitation of permission based detection method, it checks whether an application sends SMS or collects sensitive information like device ID, phone number, serial number of SIM card, and location of the device. In specifically, sendTextMessage() is checked for sending SMS, and getDeviceId(), getLine1Number(), getSimSerialNumber(), and getLastKnownLocation() are checked for collecting sensitive information. Collecting more than two kinds of the information is considered sufficiently suspicious.

### 3.2.4 Similarity scoring and malware classification

Malware applications detected by our malware detector are fed into the malware classifier module. Suspicious API strings, malicious commands, and critical permissions are used for similarity scoring. These features are chosen because they are highly related to the behavior of malware.

Similarity score between two malware is computed by using each similarity $S_i$ of suspicious API string, malicious commands, and critical permissions. The similarity score $SS$ can be written by

$$SS = \sum_i w_i \cdot S_i$$

where $w_i$ is a weight of relevant similarity, and $\sum_i w_i = 1$. For all $i$, $w_i = 1/3$ to set up same weights to three similarities.

All suspicious API are substituted by a matched character. The suspicious API string of an application, which is made of substituted characters in parsed order, reflects the calling sequence of APIs. The similarity between two strings is calculated by using Needleman-Wunsch algorithm that finds the best sharing alignment of two sequences.

The similarity of malicious commands is measured by applying the Jaccard coefficient. The Jaccard coefficient computes the number of elements in the intersection divided by the number of elements in the union. The order of the malicious commands is not under consideration.

The critical permissions are compared by using the Levenshtein distance. This metric calculates a minimum number of character edits to make two strings same. It is meaningless to consider the order of permissions, so we applied the Levenshtein distance after sorting the strings. A value of similarity is calculated as the number of edits over the maximum length of two strings. Each similarity of requested permissions and API-related permissions is computed, and the average of two values is used for the similarity of critical permissions.

Our classifier module makes groups of similar malware by comparing between a malicious application's signature and each group's signature. A signature is a set of suspicious API string, malicious commands, and requested/API-related permissions of a malicious application. In the case of a group, it is same as the signature of the first application included in the group. A sample loaded into the system is treated as follows:

1. A similarity score is computed between signatures of the sample and the existing group. The score is computed with all existing groups, each.
2. The highest value of similarity scores, max(SS), is chosen.
3. The chosen value is compared with similarity threshold $T_S$. If max(SS) ≥ $T_S$, the sample is included in the corresponding group. If max(SS) < $T_S$, the sample results in a new group, and the signature of it represents the group's signature.

## 4      Experiment Results and Discussion

We implemented our system developed by Python 2.7 using the algorithms we proposed. It is tested on collected dataset mentioned at the beginning of section 3.2. We applied 5-fold cross-validation to test our system. Samples are randomly divided into five equal size subsamples. Only one subsample is used as test data, and it repeats for five times to check for all each subsample. The accuracy is computed by averaging the five results.

### 4.1     Extracting elements

An APK, which is an Android package, is a compressed file that contains META-INF, lib, res, and assets directories and AndroidManifest.xml, classes.dex, and resources.arsc files. Serial number is extracted from META-INF directory that includes information of the certificate. From AndroidManifest.xml, the application name, requested permission, component and intent can be collected. Classes.dex file is disassembled to obtain a smali code, which is a Dalvik virtual machine code. By following the codes used by components, the system extracts suspicious API, malicious command, and API-related permission. These elements are saved in a database that provides necessary data to detector module and classifier module.

### 4.2     Serial number distribution

The distribution of serial numbers between benign samples and malware samples shows clear difference. Serial numbers found in only one application comprise the most parts of total serial numbers found in the benign dataset. The frequency of serial numbers used in benign applications decreases rapidly when number of application signed increases. In malware sample dataset, serial numbers that signs more than one application comprise larger proportion than those found in benign samples. Indeed, the average number of application signed was 1.65 in benign dataset (which includes 51,179 samples), and 7.32 in malware dataset (which includes 4,554 samples).

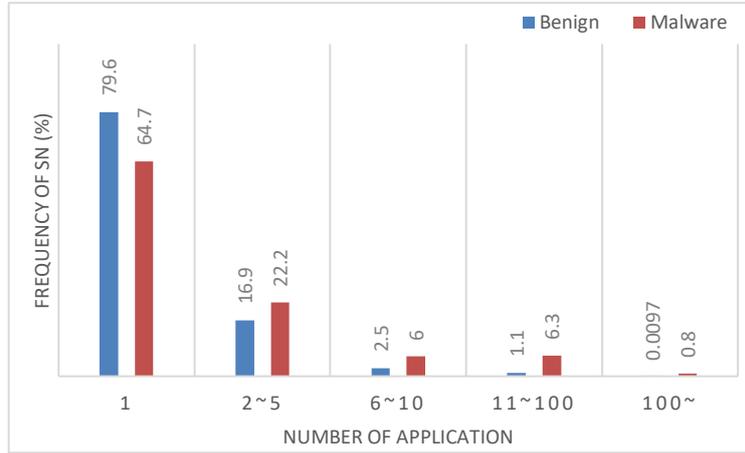

**Fig. 3.** Frequency of serial number found in sample dataset

As we mentioned when deciding the serial number blacklist in section 3.2.1, we considered the number of malware families is more important than the number of applications signed in the sense of maliciousness of a serial number. We analyzed how many malware families or variants are observed from each serial number. Table 3 shows the result. '93:6e:ac:be:07:f2:01:df' and 'b3:99:80:86:d0:56:cf:fa' are deleted from the set, since these numbers can't indicate specific malware creator. They are standard test keys for native applications which are built in a device or an emulator. The two keys, namely the test key and the platform key, are used for signing in order to develop the native applications. In 620 serial numbers, there were 484 numbers that generate only one kind of families or variants. Rest of 136 serial numbers created multiple malware families or variants. Malware dataset shows a significant trend of certificates sharing. It implies that malicious apps may have produced by professional malware creators.

**Table 3.** Number of serial numbers according to number of number of malware families

| Number of malware family (including variants) | Number of SN (except test keys) |
|---|---|
| 1 | 484 |
| 2 | **107** |
| 3 | **13** |
| 4 | **12** |
| 5+ | **4** |
| Total | 620 |

### 4.3 Detection and classification results

The system was configured with the threshold values $T_L = 1$ and $T_S = 0.7$. It is reasonable to set $T_L$ as 1 since this implies that the likelihood of malware is bigger than the likelihood of benign application. Also, we assumed that 0.7 for $T_S$ is sufficiently high to determine two signatures are similar.

423 benign applications, corresponding to 0.83% of all benign application, were detected as malware. A majority of false positives were detected by the serial number blacklist. On the other hand, the false negatives, measured as malicious samples predicted as benign, were only nine. From the point of view of malware analyst, it is important to have low false negatives, although increased false positives may be annoying. The rule may be a bit 'loose' for accurate detection. However, this substantially reduces false negatives. To find out why some of the benign applications were determined as malicious in our system, we checked the 423 false positives using descriptions from VirusTotal [22], a popular malware scanning site. Interestingly, 329 applications were reported as malware. It implies that applications downloaded from GooglePlay are not perfectly reliable as benign. The descriptions of false positives were applied to offset the classification result.

**Table 4.** Confusion matrix of malware detection results

| Category | | Predicted Class | |
|---|---|---|---|
| | | *Malicious* | *Benign* |
| **Actual Class** | *Malicious* | 4,545 (TP) | 9 (FN) |
| | *Benign* | 423 (FP) | 50,756 (TN) |

The classification accuracy of each category is analyzed in Table 5. Accuracy in the table is defined as correctly classified samples in the category divided by the number of all samples the category contains. The average of malware classification accuracy is 90%, and the average of total is 98%. Each test set of cross-validation, which contains 11,146 or 11,147 samples, was divided into 65 groups by classifier module on average. About 30 groups of them included less than 10 samples. According to the category comprising the maximum portion of group, samples of different category are false positives.

Applications in categories like Adwo, Boxer, FakeApp, FakeBattScar, Smshider, and SMStado were classified with high accuracy. However, performance of classifying sample sets like DroidDream, PremiumSMS, and SmsSend were low. The main reason was an overlap of malicious behavior. For example, SmsSend sample which is another version of OpFake according to the analysis of F-Secure, often classified as OpFake or Boxer because sending a message is a common function of them.

We conducted an experiment to estimate the detection accuracy without using the SN blacklist to see the effectiveness of the blacklist. Without SN blacklist, the detection accuracy was 98%, which shows difference of only 1% in the result of the detection with the full SN blacklist. Analyzing time (except parsing and classification process) is increased from 400 seconds to 579 seconds. Consequently, this result shows that the well-collected SN blacklist mainly boost the speed of malware detection (30.9% faster) with slightly enhanced accuracy.

**Table 5.** Classification results for each category

| Category | | Number of Samples | Accuracy |
|---|---|---|---|
| *Malware* | *AdWo* | 1,910 | **0.97** |
| | *AirPush* | 243 | 0.76 |
| | *Boxer* | 755 | **0.99** |
| | *Counterclank* | 68 | 0.68 |
| | *DroidDream* | 14 | 0.57 |
| | *DroidKungFu* | 24 | 0.79 |
| | *FakeApp* | 17 | **0.94** |
| | *FakeBattScar* | 82 | **1.00** |
| | *FakeInst* | 709 | 0.88 |
| | *FakeNotify* | 82 | 0.82 |
| | *Gappusin* | 153 | 0.65 |
| | *GinMaster* | 115 | 0.72 |
| | *Kmin* | 48 | 0.88 |
| | *OpFake* | 130 | 0.65 |
| | *PremiumSMS* | 25 | 0.44 |
| | *Ropin* | 64 | 0.66 |
| | *Smshider* | 17 | **1.00** |
| | *SmsReg* | 14 | 0.64 |
| | *SmsSend* | 10 | 0.00 |
| | *SMStado* | 74 | **0.97** |
| *Average (malware)* | | 4,554 | **0.90** |
| *Benign application* | | 51,179 | **0.99** |
| *Average (total)* | | 55,733 | **0.98** |

### 4.4 Performance evaluation

To demonstrate the performance, our system was compared with other systems: Andro-profiler [14] and Crowdroid [11]. They are classification systems using system call based on dynamic analysis. The experiment was conducted with samples used in the study of Andro-profiler. They used 709 malware and 350 benign applications. The performance was compared focusing on classification accuracy and speed. The number next to malware families implies the sample size.

Our system detects and classifies malware much better than Crowdroid, though it has slightly lower accuracy than Andro-profiler. Still, the system was implemented on larger dataset including more kinds of malware families, and it was carried out successfully with Andro-profiler's dataset at the same time.

Also, we checked the processing time of our system and Andro-profiler. The testing environment was set as Intel(R) Xeon(R) X5660 with 4GB RAM and Windows 7 Enterprise K operating system. To run Andro-profiler, the system takes 55 seconds/MB to analyze malware excluding setting time of an emulator (an added overhead). Our system performed at a speed of 72 seconds/MB. Considering dynamic analysis needs time for booting emulator between testing each sample, our system spends reasonable time for analysis.

Table 6. Comparing classification accuracy with Andro-profiler and Crowdroid

| Category | | Proposed system | Andro-profiler | Crowdroid |
|---|---|---|---|---|
| *Malware* | *AdWo(401)* | **1.00** | 1.00 | 0.54 |
| | *AirPush(60)* | **0.98** | 0.95 | 0.02 |
| | *Boxer(42)* | **1.00** | 1.00 | 0.43 |
| | *FakeBattScar(51)* | **0.96** | 1.00 | 0.18 |
| | *FakeNotify(59)* | **1.00** | 1.00 | 0.80 |
| | *GinMaster(96)* | **0.99** | 1.00 | 0.11 |
| *Benign application(350)* | | **0.88** | 0.97 | 0.35 |
| *Average* | | **0.96** | 0.99 | 0.35 |

## 5   Conclusion

In this paper, we proposed an Android malware detection and classification system based on static analysis by using serial number information from the certificate as a feature. It mainly checks a serial number, checks suspicious behavior of SMS hiding, detects the malicious system commands in the code, and analyzes the suspicious permission requests. As a result, our system can detect malware with 98% of accuracy. The classifier module can classify the 20 kinds of malware families with 90% accuracy. Additionally, the system can help analysts to react efficiently from Android malware's threats by detecting and classifying with high accuracy in a reasonable time (72 seconds/MB).

In the future, we will enhance our system by adding dynamic analysis functionality to overcome the general drawback of static analysis based system.

**Acknowledgement.** This research was supported by the MSIP (Ministry of Science, ICT and Future Planning), Korea, under the ITRC (Information Technology Research Center) support program (NIPA-2014-H0301-14-1004) supervised by the NIPA (National IT Industry Promotion Agency). This research was also supported by Basic Science Research Program through the National Research Foundation of Korea (NRF) funded by the Ministry of Science, ICT & Future Planning (2014R1A1A1006228).